\documentclass[prb,aps,showpacs,twocolumn,preprintnumbers,amsmath,amssymb,superscriptaddress]{revtex4-1}
\usepackage{amsmath,amssymb,amsfonts}
\usepackage{graphicx}
\usepackage[colorlinks=True,linkcolor=red,citecolor=blue,urlcolor=blue]{hyperref}
\usepackage{xcolor}
\usepackage{chngcntr}
\usepackage{braket}
\usepackage{hyperref}
\usepackage{nicefrac}
\usepackage{bm}
\usepackage{multirow}
\usepackage{mathtools}

\usepackage{ulem}

\def\ket#1{|#1\rangle }
\def\bra#1{\langle #1 |}

\def\ket#1{|#1\rangle }
\def\bra#1{\langle #1 |}

\begin{document}
\bibliographystyle{apsrev4-1}
\title{Band structures of (NbSe$_4$)$_3$I and (TaSe$_4$)$_3$I: Reconciling transport, optics and ARPES} 

\author{Irián Sánchez-Ramírez}
\affiliation{Donostia International Physics Center, P. Manuel de Lardizabal 4, 20018 Donostia-San Sebastian, Spain}
\affiliation{Departamento de Fisica de Materiales, Facultad de Ciencias Quimicas, Universidad del Pais Vasco (UPV-EHU), P. Manuel de Lardizabal 3, 20018 Donostia-San Sebastian, Spain}
\author{Maia G. Vergniory}
\affiliation{Donostia International Physics Center, P. Manuel de Lardizabal 4, 20018 Donostia-San Sebastian, Spain}
\affiliation{Max Planck Institute for Chemical Physics of Solids, 01187 Dresden, Germany}

\author{Claudia Felser}
\affiliation{Max Planck Institute for Chemical Physics of Solids, 01187 Dresden, Germany}

\author{Fernando de Juan}
\affiliation{Donostia International Physics Center, P. Manuel de Lardizabal 4, 20018 Donostia-San Sebastian, Spain}
\affiliation{Departamento de Fisica de Materiales, Facultad de Ciencias Quimicas, Universidad del Pais Vasco (UPV-EHU), P. Manuel de Lardizabal 3, 20018 Donostia-San Sebastian, Spain}
\affiliation{IKERBASQUE, Basque Foundation for Science, Maria Diaz de Haro 3, 48013 Bilbao, Spain}

\date{\today}
%%%%%%%%%%%%%%%%%%%%%%%%%%%%%%%%%%%%%%%%%%%%%%%%%%%%%%%%%%%%%%%%%%%%%%%%%%%%%
\begin{abstract}
Among the quasi one-dimensional transition metal tetrachalcogenides (MSe$_4$)$_n$I (M=Nb,Ta), the $n=3$ compounds are the only ones not displaying charge density waves. Instead, they show structural transitions with puzzling transport behavior. They are semiconductors at the lowest temperatures, but their transport gaps are significantly smaller than those inferred from ARPES and optical conductivity. Recently, a metallic polytype of (TaSe$_4$)$_3$I has been found with ferromagnetism and superconductivity coexisting at low temperature, in contrast to previous reports. In this work we present detailed ab-initio and tight binding band structure calculations for the different (MSe$_4$)$_n$I reported structures. We obtain good agreement with the observed transport gaps, and explain how ARPES and optics experiments effectively probe a gap between different bands due to an approximate translation symmetry, solving the controversy. Finally, we show how small extrinsic hole doping can tune the Fermi level through a Van Hove singularity in (TaSe$_4$)$_3$I and discuss the implications for magnetism and superconductivity.
\end{abstract}
\maketitle

\section{Introduction} 

The series of transition metal tetrachalcogenides (MSe$_4$)$_n$I with metals M=Nb,Ta have been long recognized  as ideal quasi-1D materials to study charge density wave (CDW) transitions~\cite{Meerschaut77,Gressier85,Gressier85b}. These compounds are made of weakly coupled MSe$_4$ chains, each of them hosting a single one-dimensional band with fractional filling $\delta = (n-1)/2n$, with reported structures occuring at $n=2,3,\tfrac{10}{3}$ \textit{i.e.} $\delta = 0.25, 0.33, 0.35$ respectively. These fractionally filled quasi-1D bands are prone to CDW instabilities which have been extensively studied \cite{Meerschaut77,Gressier85,Gressier85b,Izumi84,Izumi84b,Taguchi86,Smontara86,Izumi87,Sekine87,Sekine88,SaintPaul88,Zelezny89,Staresinic06,Dominiko16}.

In the better known Nb compounds, the $n=2$ and $n=\tfrac{10}{3}$ variants show clear incommensurate CDW transitions arising from a metallic parent state. On the other hand, $n=3$ compounds stand out as the only ones without a CDW. Instead, they display a structural transition and the activated transport of a semiconductor. While such behaviour is believed to originate from the commensurability of the unit cell and filling for $n=3$, several puzzles have precluded a consistent understanding of these materials in terms of band theory: Often more than one structural transition is observed~\cite{Gressier85,Izumi87,Sekine87,Sekine88}, and the measured transport gaps are quite variable~\cite{Gressier84,Taguchi86,Izumi84,Colluza93,Taguchi86,Taguchi87}. In addition, ARPES~\cite{Vescoli00,Grioni01} and optical conductivity~\cite{Vescoli00} experiments reported a gap that was much larger than any of those measured in transport. 

For the isoelectronic Ta compounds, only $n=2,3$ variants are reported in the literature~\cite{Gressier84,Roucau84,Gressier84b}. While (TaSe$_4$)$_2$I has seen a renewed interest \cite{Tournier13} in the context of axionic CDW in Weyl semimetals~\cite{Gooth19,Shi21}, knowledge about (TaSe$_4$)$_3$I is rather scarce, as it was assumed to behave mostly like its Nb counterpart~\cite{Gressier84,Roucau84,Gressier84b}. In a recent development, however, a non-centrosymmetric polytype of (TaSe$_4$)$_3$I has been found to be metallic at room temperature, with a ferromagnetic transition at 8 K, and superconductivity coexisting with ferromagnetism at $T_c = 3$ K~\cite{Bera21}. This coexistence is unusual on its own~\cite{Aoki01,Rosner01}, but it is even more surprising given the isoelectronic Nb compound semiconducting behaviour. 

The conflicting transport and ARPES results for (NbSe$_4$)$_3$I, along with the unexpected low temperature behavior of (TaSe$_4$)$_3$I reveal that an understanding of these materials by means of band theory is currently lacking. In this work, we combine ab-initio calculations of all known structures for both compounds with a tight-binding analysis to explain the fundamental electronic properties of these systems. All phases are found to be semiconducting, with structures showing a trimerization distortion of the MSe$_4$ units and with gaps that can be directly correlated with the amount of trimerization. We discuss how these gaps compare with transport observations and find in general good agreement (except for the new metallic polytype, which is probably unintentionally doped). However, we also show that while the gap is formally direct and located at $\Gamma$, there is an approximate in-plane translational symmetry leading to a negligible spectral weight for the valence band edge, so ARPES and optical conductivity actually probe higher bands and lead to a misidentification of the true gap. Finally, we identify a Van Hove singularity in the valence band edge in the new (TaSe$_4$)$_3$I polytype which crosses the Fermi level for a very small amount of hole doping, and we argue that this might be relevant for the observed magnetism and superconductivity. 

\section{Phenomenology overview} \label{Phenomenology}

For the $n=3$ compounds, (NbSe$_4$)$_3$I is the best studied one. Its high temperature phase has a tetragonal, centrosymmetric crystal structure with space group (SG) $P4/mnc$ (No. 128, point group $D_{4h}$). As temperature is lowered, it undergoes a structural phase transition at $T_{c1} \sim 274-280$ K with the resulting space group $P \bar{4} 2_1 c$ (No. 114, point group $D_{2d}$). A second structural transition to a phase with SG $P\bar{4}$  (No. 81, point group $S_4$) at $T_{c2}\sim 90$ K has also been reported~\cite{Gressier85,Izumi87,Sekine87,Sekine88}, but not in all experiments. In this work, we will refer to these structures only by their point group. The pattern of symmetry breaking is thus $D_{4h}\rightarrow D_{2d} \rightarrow S_4$. 

Resistivity measurements are generally consistent with semiconducting behavior with activated resistivity $\rho \propto e^{E_g/2k_B T}$ and a gap $E_g$ that appears to change through the phase transitions. Above $T_{c1}$, values of $E_g =$ 190-220 meV \cite{Gressier84,Dominiko16} were reported. A resistivity kink was always observed at $T_{c1}$, and below it two types of samples were reported to exist~\cite{Gressier84}, initially indistinguishable by structural measurements~\cite{Gressier84,Taguchi86}. In type I samples $E_g$ is reduced to values in the range 20-70 meV \cite{Gressier84,Taguchi86,Izumi84,Colluza93} all the way to the lowest temperatures measured. In type II samples, a broader kink is observed which leads to low temperature gaps of 110-130 meV \cite{Gressier84,Dominiko16}. However, in some samples $\partial \ln \rho/\partial T^{-1}$ never flattens to a constant $E_g$ value, but rather continuously decreases after a maximum~\cite{Taguchi86,Taguchi87}, challenging the view of standard activated transport. Differing observations also include a report of $E_g =$ 97 meV for $T>T_{c1}$ and 222 meV for $T<<T_{c1}$\cite{Sekine88}, or even a an abrupt change at 180 K from 345 meV to 22 meV\cite{Smontara86}. Another characteristic feature of type II samples is that switching to a state of lower resistance can be induced at high currents and low temperatures~\cite{Taguchi87}. This behavior disappears at 140 K, and is completely absent in type I crystals.

A partial solution to these transport puzzles was offered in Ref. \cite{Dominiko16} which reported that the second structural transition was only observed in samples assigned to type II in transport. The overall suggested picture would then be that all three structures are semiconductors: the initial gap of 190-220 meV for the $D_{4h}$ structure is reduced to 20-70 the $D_{2d}$ structure, and for samples with the second transition it increases again in the $S_4$ structure to 110-130. While outliers to this picture do exist, we consider this to be the average behaviour to compare with our calculations. 

Early ARPES experiments at 300 K also attempted to determine the spectral gap~\cite{Vescoli00,Grioni01}. The valence band maximum was found to be at 750 meV below $E_F$, providing a lower bound on the gap which is much larger than the one obtained from every transport experiment. Optical conductivity also showed a raising edge at $\omega \sim$ 500 meV \cite{Vescoli00}, again too large compared to transport gaps. The fact that the gap derived from ARPES and optical conductivity is much larger than the ones obtained from transport remains an unsolved problem to date, precluding any consistent band-structure understanding of these materials.  

Finally, much less is known about (TaSe$_4$)$_3$I~\cite{Gressier84,Roucau84,Gressier84b}, which has generally been assumed to behave like its Nb counterpart. The $D_{4h} \rightarrow D_{2d}$ structural phase transition was measured at at 200 K~\cite{Roucau84}, but the $S_4$ phase has not been reported. Very recently, the $D_{2d}$ was reported to be metallic with a resistance kink-plateau at 150 K, a ferromagnetic transition at 8 K, and a superconducting one at $T_c = 3$ K \cite{Bera21}, in stark contrast to previous observations. The coexistence of ferromagnetism and superconductivity is a rarely reported phenomenon~\cite{Aoki01,Rosner01}, and it is often taken as a hint that pairing could be in a unconventional odd-parity triplet channel. While experiments on superconductivity are at a very early stage, a band structure understanding of the basic properties of (TaSe$_4$)$_3$I is clearly needed as a starting point to understand these unusual behavior.

\begin{figure}[t]
    %\vspace{-1cm}
    %\hspace{-1cm}
    \centering
    \includegraphics[width=0.5\textwidth]{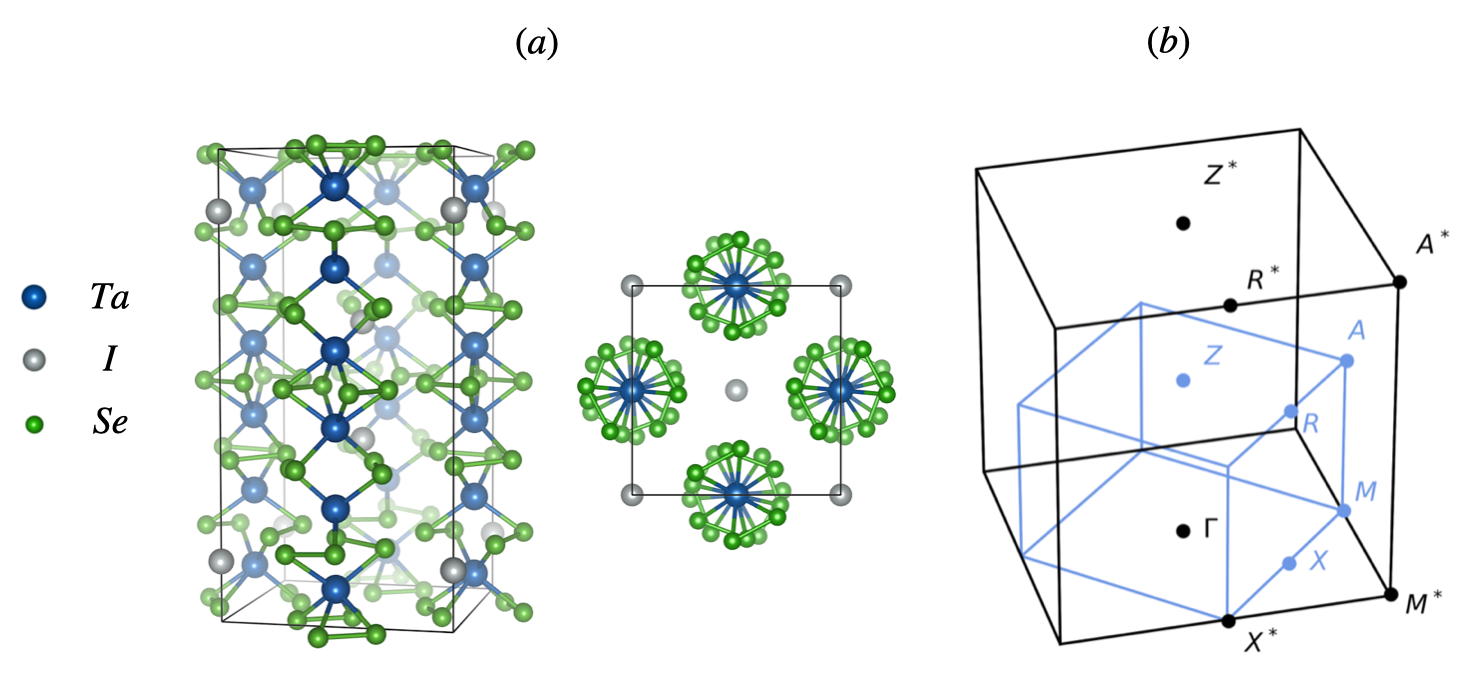}
    \vspace{-0.5cm}
    \caption{(a) Atomic structure of (TaSe$_4$)$_3$I in the $D_{4h}$ phase and (b) Brillouin zones for the atomic structure of in (a) shown in blue, and for the approximate effective structure described in the text which contains one formula unit}
    \label{Fig:structure}
\end{figure}

\begin{figure*}[t]
    \hspace{-0.5cm}
    \centering
    \includegraphics[width=0.95\textwidth]{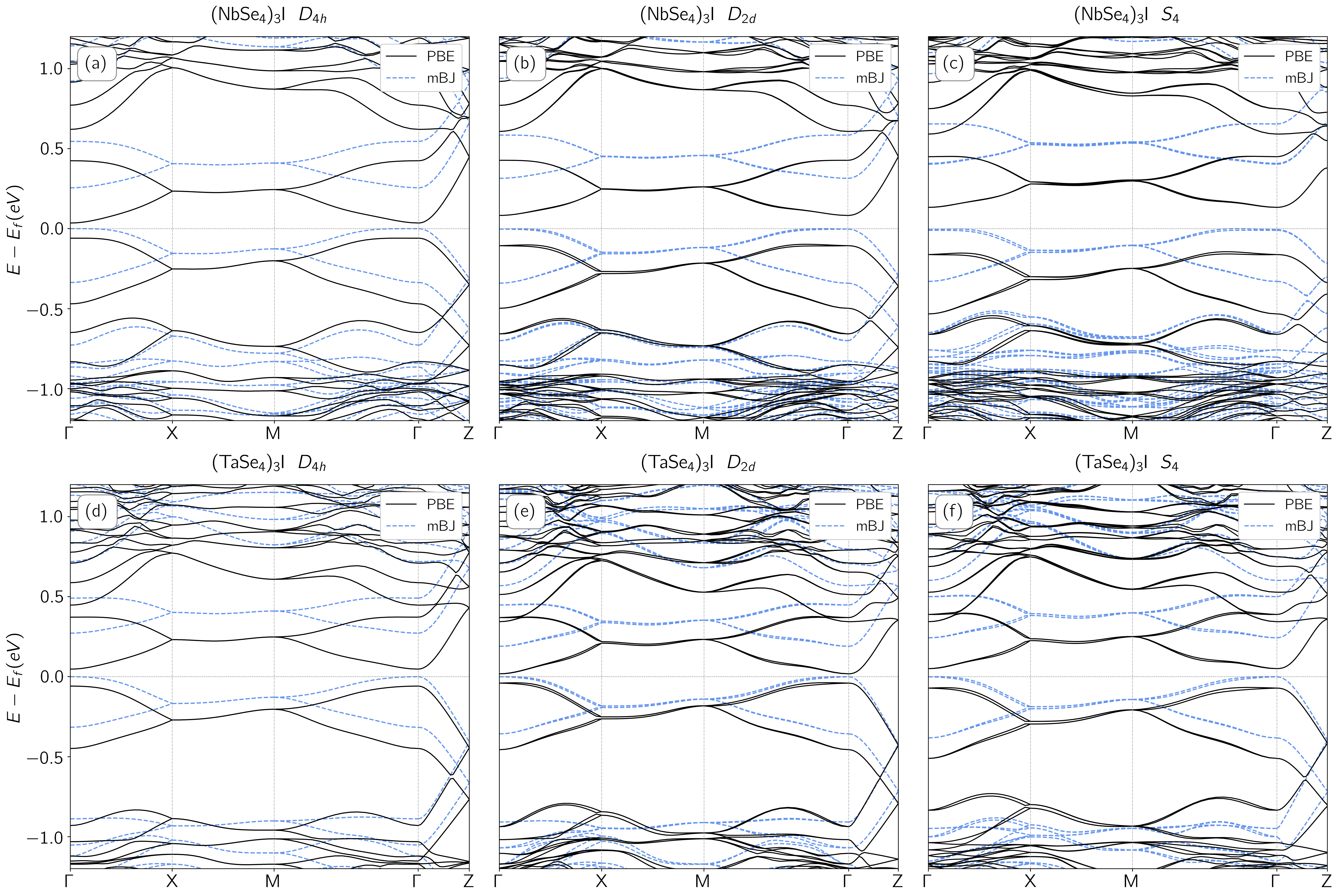}
    \caption{Ab-initio bandstructures for the (MSe$_4$)$_3$I compounds in the three different structures $D_{4h}$, $D_{2d}$ and $S_4$ from left to right. Top row (a-c) shows (NbSe$_4$)$_3$I, bottom row (d-f) shows (TaSe$_4$)$_3$I. Solid back lines were obtained using PBE pseudopotentials and dashed blue lines were obtained using mBJ hybrid potentials.}
    \label{Fig:mbj_nsi}
\end{figure*}

\iffalse

\begin{figure*}[h!]
    \hspace{-0.5cm}
    \centering
    \includegraphics[width=0.95\textwidth]{nsi_horiz.png}
    \caption{Ab-initio bands for (NbSe$_4$)$_3$I SG (a) 128, (b) 114 and (c) 81 structures. Solid back lines were obtained using PBE pseudopotentials and dashed blue lines were obtained using mBJ hybrid potentials.}
    \label{Fig:mbj_nsi}
\end{figure*}

\begin{figure*}[h!]
    \hspace{-0.5cm}
    \centering
    \includegraphics[width=0.95\textwidth]{tsi_horiz.png}
    \caption{Ab-initio bands for (TaSe$_4$)$_3$I SG (a) 128, (b) 114 and (c) 81 structures. Solid back lines were obtained using PBE pseudopotentials and dashed blue lines were obtained using mBJ hybrid potentials.}
    \label{Fig:mbj_tsi}
\end{figure*}

\fi

\begin{figure}[h]
    \centering
    \hspace*{-0.8cm}
    \includegraphics[width=0.55\textwidth]{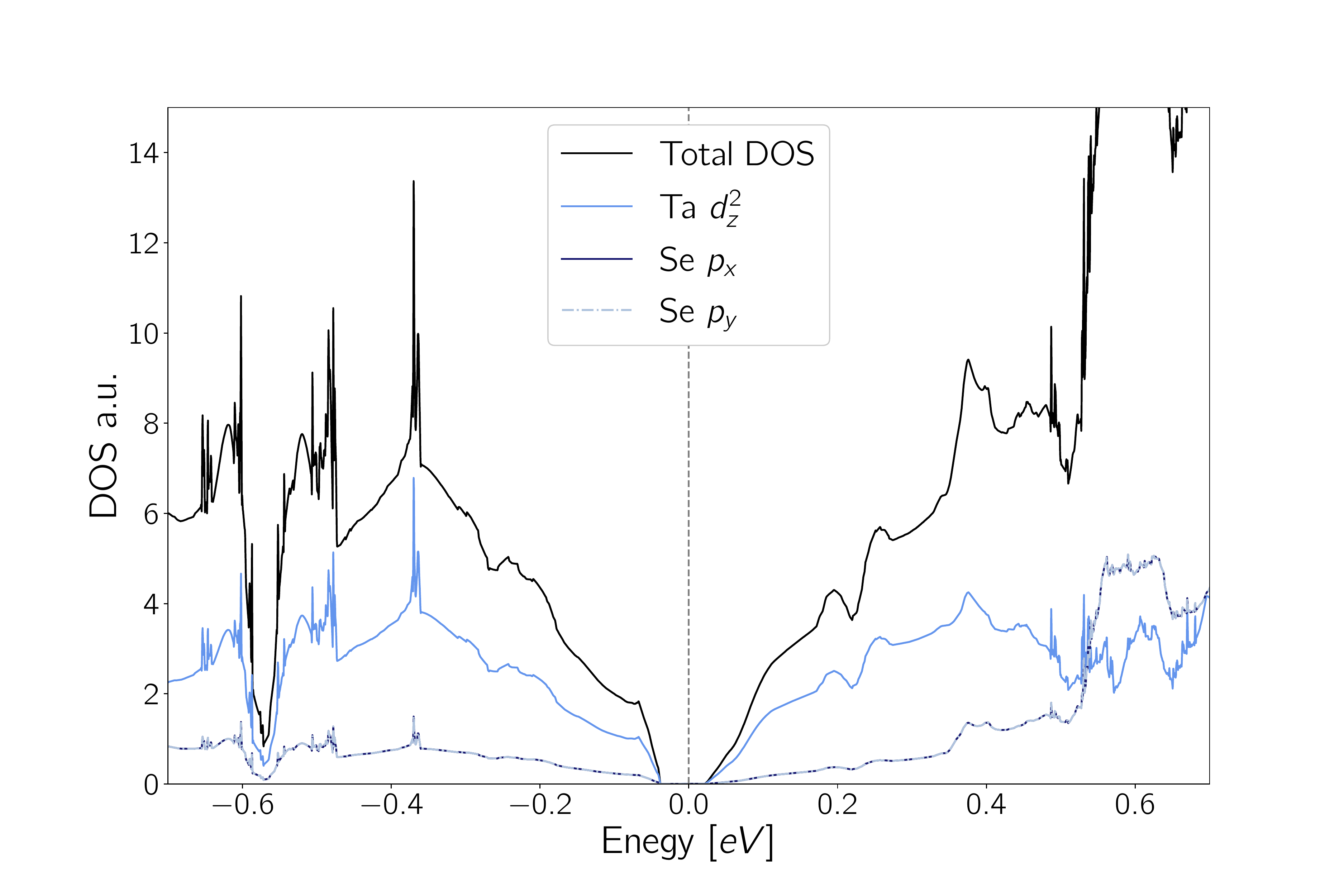}
    \caption{Orbital resolved density of states for (TaSe$_4$)$_3$I in the $D_{2d}$ phase computed within the PBE parametrization. This density of states corresponds to the bands shown in Fig. \ref{Fig:mbj_nsi} (e). The bands near the Fermi level have dominant $d_{z^2}$ orbital character on the Ta sites. This is true for all computed band structures (not shown).}
    \label{Fig:pDOS_bigTB}
\end{figure}

\section{Methods}

With the aim of explaining the previously discussed phenomenology, first-principles density functional theory (DFT) band structure calculations for the different reported structures were performed using VASP\cite{VASP1,VASP2} \textit{v.}$6.2.1$ with projector-augmented wave pseudopotentials using two approximations: the generalized gradient approximation with Perdew Burke Ernzerhof parametrization (PBE)  \cite{Perdew96} and the modified Becke-Johnson method (mBJ) \cite{Tran2009}. The mBJ method was chosen in order to obtain a more accurate description  of the bandgaps of the studied structures, in the interest of unraveling the possible metallic behaviour of (TaSe$_4$)$_3$I in the $D_{2d}$ phase.

Fig. \ref{Fig:structure} shows the crystal structure of (MSe$_4$)$_3$I compounds in the $D_{4h}$ phase. The MSe$_4$ units form one dimensional chains, where each M atom is sandwiched between Se$_4$ rectangular units and separated by I$^-$ ions. Each Se$_4$ adjacent rectangular unit is rotated around $45^{\circ}$ and modulated by small distortions on Se positions. The typical unit cell for these compounds is tetragonal ($a=b\neq c$) and contains two metallic chains. The particular case of the Fig. \ref{Fig:structure} is (TaSe$_4$)$_3$I in $D_{4h}$ phase, with $a=b=9.719~\text{\AA}$ and $c=19.363~\text{\AA}$. The difference between phases \textit{i.e} $D_{4h}$, $D_{2d}$ and $S_4$; is given by the relative distance between M atoms and small modulations in the Se positions.

For (NbSe$_4$)$_3$I, a detailed structural characterization is available for the three considered phases $D_{4h}$\cite{Meerschaut77}, $D_{2d}$\cite{Gressier85}, and $S_4$\cite{Izumi87}. Therefore, all DFT calculations for these compounds were performed using the experimentally-measured lattice parameters. Self-consistent calculations considering spin-orbit coupling were found to be well converged for a kinetic energy cutoff of $520 \;\text{eV}$ and a $9\times9\times5$ k-mesh sampling. Conversely, in the case of (TaSe$_4$)$_3$I, no detailed characterization exists. Given the similarity with the Nb compounds~\cite{Gressier85}, we obtained the structures of the Ta compounds performing a structural relaxation starting with the positions of the experimentally-measured Nb-based compounds\cite{Izumi84,Jain2013}.  The relaxation calculations were performed using a conjugate gradient algorithm\cite{Shampine87} and were found to be well converged for a kinetic energy cutoff of $520 \;\text{eV}$ and a $11\times11\times5$ k-mesh sampling. For phases $D_{4h}$ and $D_{2d}$, the starting point were the pre-relaxed  structures in Ref.\cite{Jain2013} while, for phase $S_4$, a direct relaxation with the conjugate gradient algorithm was performed from the original positions of the isoelectronic Nb compound in the same phase since no pre-relaxed data was available. This strategy prioritizes the conservation of the space-group symmetries from the original Nb structures. Self-consistent calculations considering spin-orbit coupling were found to be well converged for a $520\:\text{eV}$ kinetic energy cutoff and a $11\times11\times5$ k-mesh. Density of states calculations were performed using a $15\times15\times9$ k-mesh and a energy resolution of $0.7 \;\text{meV}$. 

\begin{table*}[t]
\scalebox{0.9}{
\begin{tabular}{|c|c|c|c|c|c|c|c|}
\hline & Structure & $a \;($\AA$)$ & $c \;($\AA$)$ & $\left(\frac{1}{6}-\frac{d_{M_i-M_{i+1}}}{c}\right)^2\;(\times10^{-4})$ & $ \Delta d$ & Gap PBE ($eV$) & Gap mBJ ($eV$)\\
\hline
    \multirow{4}{*}{Nb} & $D_{4h}$ & 9.4891 & 19.1323 & 0.11 - 0.11 - 0.46 - 0.11 -  0.11 - 0.43 & 1.15 & 0.09 & 0.26 \\
    & $D_{2d}$ & 9.4500 & 19.0799 & 0.00 - 0.47 - 0.39 - 0.00 - 0.47 - 0.39  & 1.32 & 0.18 & 0.31 \\
    & $S_4$ & 9.4365 & 19.0461 & 1.07 - 0.17 - 0.12 - 0.58 - 0.00 - 1.01   & 1.71 & 0.28 & 0.40 \\
    \hline
    \multirow{4}{*}{Ta} 
    & $D_{4h}$  & 9.7192 & 19.3626 & 0.09 - 0.09 - 0.37 - 0.09 - 0.09 - 0.37  & 1.05 & 0.11 &  0.27\\
    & $D_{2d}$ & 9.4365 & 19.4365 & 0.23 - 0.00 - 0.26 - 0.23 - 0.00 - 0.28  & 1.00 & 0.06 & 0.19\\
    & $S_4$ & 9.4365 & 19.0461 & 0.00 - 0.39 - 0.26 - 0.01 - 0.33 - 0.30 & 1.13 & 0.12 & 0.24\\
    \hline
\end{tabular}}
\caption{Lattice parameters, metal-metal distances, distortion $\Delta d$ and bandgaps for PBE and mBJ approximations and hybrid potential for the different structures considered in the text.}
\label{distortions}
\end{table*}

\section{DFT band structures} \label{DFT}

The electronic band structures for (NbSe$_4$)$_3$I and (TaSe$_4$)$_3$I are presented in Fig.~\ref{Fig:mbj_nsi} in both PBE and mBJ approximations. The band structures for both Nb and Ta compounds are overall similar for all phases. A set of 8 low energy bands is observed in the energy window $E\in [-0.5,0.5]$~eV. This set is found closer to the valence bands in the Nb compounds compared to the Ta compounds. These bands have a dominant orbital weight coming from $d_{z^2}$ orbitals of the Ta atoms, as shown in the density of states in Fig.~\ref{Fig:pDOS_bigTB} and as explained originally by Gressier \textit{et.~al.}~ \cite{Gressier84b,Gressier85b}. The fact that these bands have dominant weight in a single Ta orbital suggests a simple tight binding model should describe these bands correctly, as shown in the following section. 

Regarding the band gaps, in the case of (NbSe$_4$)$_3$I, the mBJ method shows a greater band gaps than the PBE parametrization for the three phases. In PBE, the band gap is slightly indirect, with the valence band maximum slightly off $\Gamma$ in the $M-\Gamma$ direction and the conduction band minimum at $\Gamma$. In mBJ, the band gap is also indirect for $D_{2d}$ and $S_4$ phases, but it is direct for $D_{4h}$. Similarly, for (TaSe$_4$)$_3$I the mBJ method also shows a greater band gap than PBE for all phases. Both in mBJ and PBE approximations the band gaps are direct between $\Gamma - \Gamma$ for inversion conserving phase $D_{4h}$ and slightly indirect for inversion-broken phases $D_{2d}$ and $S_4$. 

In order to understand the different values for the obtained from DFT calculations we recall that the relative distances between Nb atoms were reported to be related to the electronic band gap in the past. In the interest of gaining insight of this statement for (MSe$_4$)$_3$I compounds we introduce the concept of \textit{distortion} $\Delta d$ 
\begin{align}
\Delta d = \sqrt{\sum_{i=1}^6 \left(\frac{1}{6} - \frac{d_{M_i-M_{i+1}}}{c}\right)^2}
\label{eq:def_dist},
\end{align}
where $d_{M_i-M_{i+1}} = \hat{d}_i$ is the distance between neighbouring M atoms along the $\vec{c}$ direction, $c=|\vec{c}|$ and the sum runs over the 6 M atoms in the (MSe$_4$)$_3$I compounds. The quantity $\Delta d$ measures how far a 6-atom chain is from being evenly spaced, and vanishes when all distances between M atoms are equal, \textit{i.e.} $\hat{d}_i=\hat{d}_j $ for all $i,j$. 
A summary of relevant structural data and distortions, along with the band gaps obtained both in the PBE parametrization and mBJ method are presented in Table \ref{distortions}. The results in the table suggest an relation between distortion, cell volume and electronic band gap. In general terms, a greater distortion leads to wider electronic band gaps. Even so, the cell volume also seems to plays a role in enhancing or reducing the band gap, since smaller cell volumes lead to smaller band gaps. The data recollected in Sec.~\ref{Phenomenology} for transport band gaps in Nb-compounds, shows the following trend: $\Delta E_{D4h} > \Delta E_{S4} > \Delta E_{D2d}$, while our data suggests that $\Delta E_{S4} > \Delta E_{D2d} > \Delta E_{D4h}$. Even though the trends are dissimilar, the magnitudes for the DFT electronic gaps are comparable to the ones obtained with transport measurements.

\section{Tight-Binding models} \label{tb}

\begin{figure*}[t]
    \centering
    \includegraphics[width=0.95\textwidth]{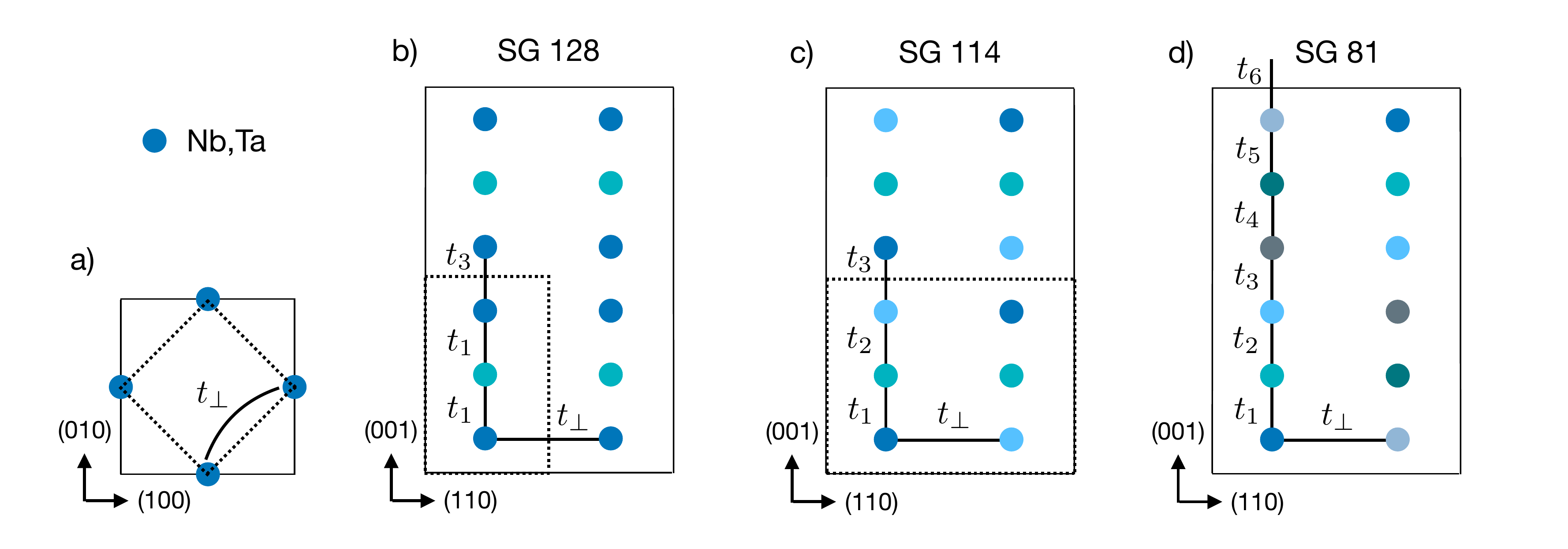}
    \caption{Minimal tight binding models for the three different space groups. Inequivalent M sites are drawn with different colors, and reduced unit cells due to approximate translation symmetries are depicted with dashed lines. a) View from the top and b) from the side for $D_{4h}$ phase. c) $D_{2d}$ phase. d) $S_4$ phase.}
    \label{Fig:TBcell}
\end{figure*}

Since the orbital-resolved density of states in Fig. \ref{Fig:pDOS_bigTB} shows that the bands near the Fermi level are dominated by M $d_{z^2}$ orbitals, we can gain a deeper understanding of the structural dependence of the band gap from a simple tight-binding model containing only such orbitals in Ta sites. The model will be constrained by the symmetries of each phase: $D_{4h}$ is generated by inversion $I$, the glide $\{m_{110}|\tfrac{1}{2}\tfrac{1}{2}\tfrac{1}{2}\}$, the rotoinversion $S_4$ and the two-fold screw $\{2_{100}|\tfrac{1}{2}\tfrac{1}{2}\tfrac{1}{2}\}$. Breaking of inversion then leads to the point group $D_{2d}$ and further breaking the glide and two-fold axis leads finally to point group $S_4$.

In the phase of higher symmetry, $D_{4h}$, there are only two non-equivalent M atoms, as shown in Fig. \ref{Fig:TBcell}a) and b). The simplest model therefore only contains two different on-site potentials $\Delta_1$ and $\Delta_2$, two intrachain hoppings $t_1$ and $t_2$ and a single interchain hopping $t_\perp$. As the figure shows, this model actually has an accidental translation symmetry, with a reduced unit cell with 3 M sites, representing a single formula unit (MSe$_4$)$_3$I (the true unit cell would have 4 formula units). The lattice parameters of this reduced unit cell are $c^* = c/2$ and $a^* = a/\sqrt{2}$, and the corresponding enlarged Brillouin Zone BZ$^*$ is shown in Fig. \ref{Fig:structure}, with high symmetry points denoted as $A^*$, $Z^*$ and so on.  The Hamiltonian for this model is 
 
\begin{align}
H = -\left(\begin{array}{ccc}
\Delta_1 + t_\perp f(k_\parallel)  & t_1 e^{i k_z c/6} & t_3 e^{-ik_z c/6} \\
t_1 e^{-i k_z c/6}     & \Delta_2+ t_\perp f(k_\parallel) & t_2 e^{i k_z c/6} \\
t_3 e^{ik_z c/6} & t_2 e^{-ik_z c/6} & \Delta_3+ t_\perp f(k_\parallel)
\end{array}   \right) \label{Eq:3bandTB}
\end{align}
where $t_1 = t_2$ and $$f(k_\parallel) = 2\left[\cos \left(\frac{(k_x+k_y)a}{2\sqrt{2}}\right)+\cos \left(\frac{(k_x-k_y)a}{2\sqrt{2}}\right)\right].$$
To account for the trimerization distortion, we set $t_2 = t -\delta /2$ and $t_3 = t + \delta$ so that the average hopping is $t$ and $\delta$ parametrizes the trimerization assuming that the hoppings change linearly with the distance between orbitals. 

The presence of the extra translation symmetries in this model is accidental, as the inclusion of further neighbor intra- and interchain hoppings would indeed require the use of the full 12 site unit cell. This further neighbor hoppings are however expected to be smaller, so that the model in Eq. \ref{Eq:3bandTB} serves as a good first approximation for the band structure. Physically, this means that the actual positions of the Se$_4$ units and I$^-$ ions has little effect on the low energy M-derived bands. 

\begin{figure*}[t]
    \centering
    \includegraphics[width=0.9\textwidth]{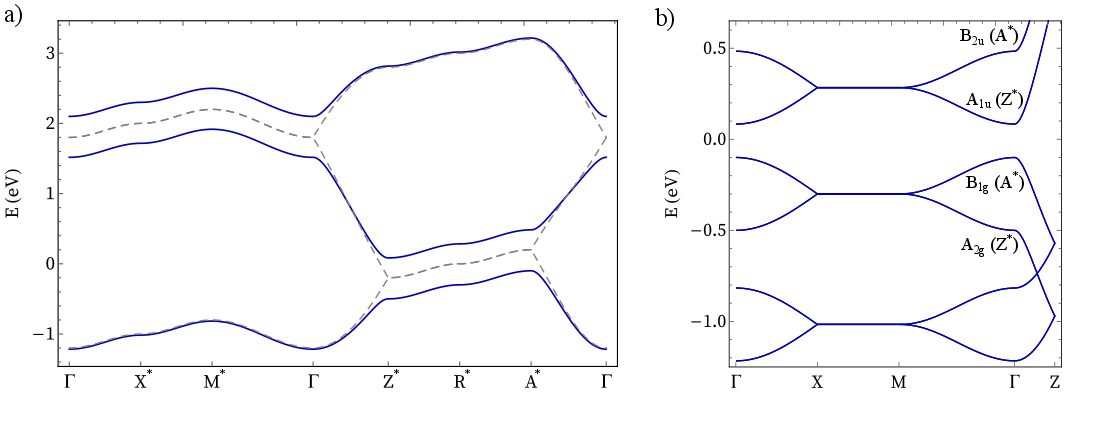}
    \caption{a) Band structure from the simplified tight-binding model in Eq. \ref{Eq:3bandTB}. Gray dashed bands represent the hypothetical structure without trimerization ($\delta=0$.) b) Closeup near the Fermi level for the band structure for the physical 12-Ta unit cell, where the four occupied states at $\Gamma$ fold from $A^*$, $R^*$, $M^*$ and $\Gamma$ in the BZ$^*$, in order of decreasing energy.}
    \label{Fig:TB}
\end{figure*}

Fig. \ref{Fig:TB} (a) shows the bands obtained from Eq. \ref{Eq:3bandTB} for  $\Delta_i=0$, $t=1$ eV, $\delta = 0.3$ eV and $t_\perp = 0.05$ eV. For comparison, we also show the same plot for $\delta=0$, and we see that a finite distortion $\delta$ opens a gap a the Fermi level. A finite value of $\Delta_i$ would further contribute to the opening of the gap and is not considered for simplicity. We also observe that the presence of small interchain hopping leads to an indirect gap, with the valence band maximum at $A^*$ and the conduction band minimum at $Z^*$. 

To compare the tight-binding model with the computed ab-initio band structures, in Fig. \ref{Fig:TB}b) we also plot the bands in the physical 12 site unit cell. Doing so we observe a backfolding of the bands, so that below the Fermi level we now have four bands at $\Gamma$, originating from $A^*$, $R^*$, $M^*$ and $\Gamma$. The general agreement of the folded bands with the ab-initio band structures in Fig. \ref{Fig:mbj_nsi} suggests that this reduced model is indeed a very good approximation. 

Another prediction of this simple model is the irreducible representations (irreps) of the different bands near the Fermi level at $\Gamma$. Depending on the signs of the hoppings we get different irreps that we can directly compare with the irreps computed from ab-initio wavefunctions.
Fig. \ref{Fig:TB}b) shows the irreps for the natural choice $t_i>0$, $t_\perp>0$, for the four nearest bands to the Fermi level (two occupied and two unoccupied), which have symmetries $(A_{2g},B_{1g},A_{1u},B_{2u})$ from the lowest to the highest energy band respectively. Using the software {\tt vasp2trace}\cite{V2T1,V2T2} we have computed the irreps of the same bands ab-initio and found exact agreement, thus confirming the correct sign of the hoppings. 
We note that our computed irreps differ from the assignment proposed in Ref.~\cite{Sekine88} which would require $t_\perp<0$ and would lead to the conduction and valence bands locating at different momenta in the unfolded model. 

The conclusion of this analysis is that, to a very good approximation, (MSe$_4$)$_3$I in the $D_{4h}$ phase behaves as if its band structure was given by the three site model, i.e. as a semiconductor with an indirect gap, where the valence band maximum occurs at $A^*$ while the conduction band minimum occurs at $Z^*$. It is only due to very small perturbations induced by further neighbor hoppings (or physically, by the structure of the Se cages and I atoms), that the backfolding is induced.

Finally, Fig \ref{Fig:TBcell}c) and d) show the corresponding models for the $D_{2d}$ and $S_4$ phases, respectively. The $D_{2d}$ structure requires a 6 M atom unit cell, which induces a folding of the unit-cell only in the plane, now requiring two chains per unit cell, while the effective $c^*$ translation is preserved. When the phase transition from $D_{4h}$ to $D_{2d}$ occurs, a folding of $A^*$ to $Z^*$ occurs, so that the band gap now becomes direct even in the simplified model. Finally, in the $S_4$ structure no effective translation remains, as all 6 M sites within a chain are inequivalent.

\section{Unfolding ab-initio and ARPES} \label{ARPES}

\begin{figure}[t]
    \centering
    \hspace{-0.5cm}
    \includegraphics[width=0.42\textwidth]{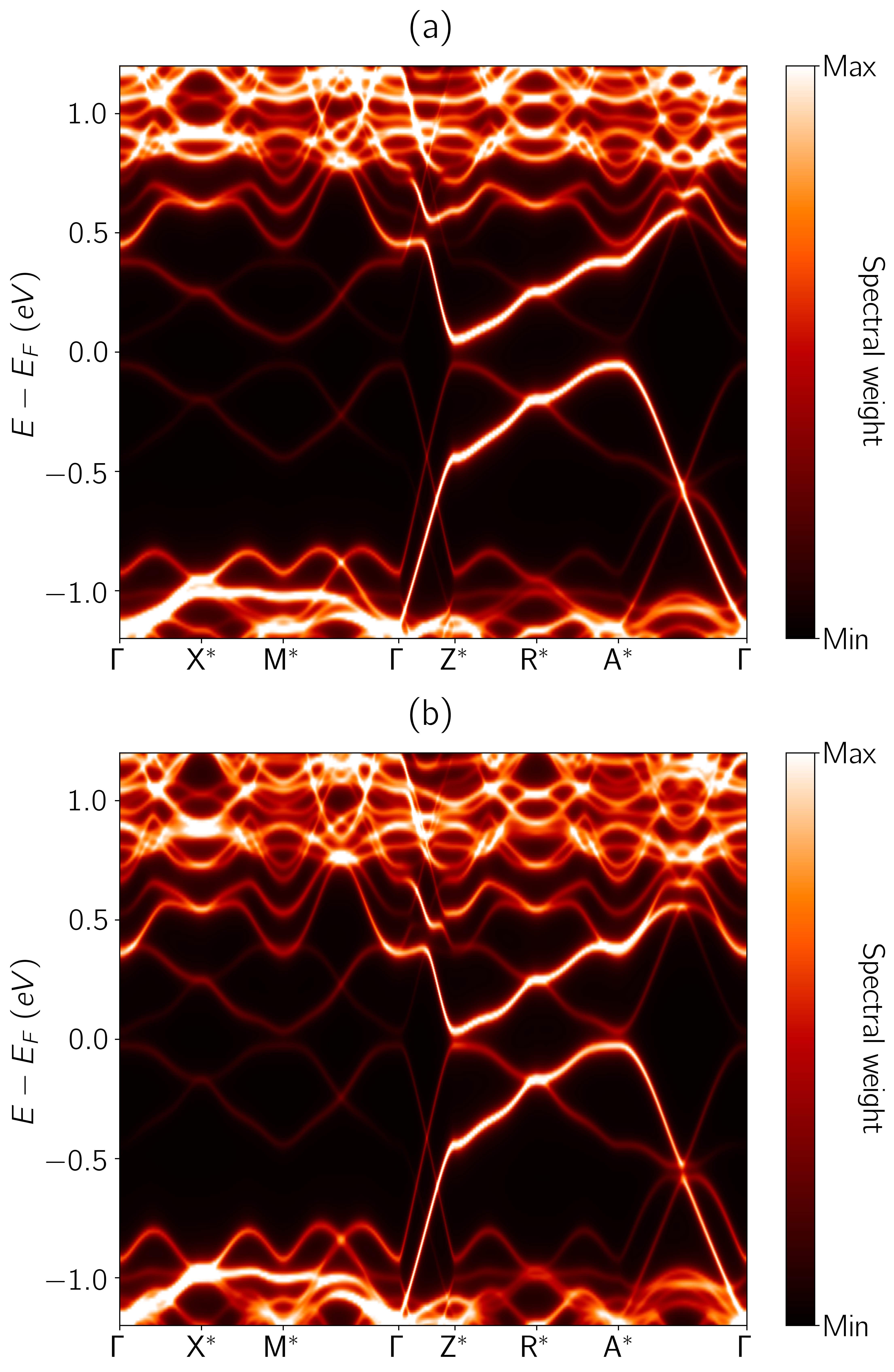}
    \caption{Unfolded band structure for the unit cell with one formula unit in (a) $D_{4h}$ and (b) $D_{2d}$ phase \cite{Popescu12}. Compare with the corresponding TB model in the same unit cell in Fig. \ref{Fig:TB} (a).}
    \label{Fig:Unfold}
\end{figure}

In order to ascertain to which extent the picture presented in Sec.~\ref{tb} is quantitatively correct, we now consider an ab-initio calculation of the unfolded spectral function, following the method of Popescu and Zunger\cite{Popescu2012} as implemented in {\tt VaspBandUnfolding}\cite{Vaspunfold}. When a crystal with a given unit cell is perturbed by a superlattice modulation, this method can be used to compute the spectral function corresponding to the original unit cell in terms of an effective band structure. In our case, the modulated structure is actually the physical crystal structure with 12 M atoms, but we interpret it as a periodically modulated crystal with an original unit cell with 3 M atoms. The unfolded band structures of the $D_{4h}$ and $D_{2d}$ phases of the Ta compound are shown on Fig.~\ref{Fig:Unfold}. 

In Fig. \ref{Fig:Unfold}a) only two bands show a significant spectral weight near the Fermi energy ($E_F$ $\pm 0.5$ eV). To great extent, this energy dispersion matches the one obtained from the 3 site tight-binding shown in Fig.~\ref{Fig:TB} and the previously identified band edges: indeed, while the band edge formally occurs at $\Gamma$ in the full calculation, it is actually backfolded from $A^*$ due to a minuscule perturbation caused by the Se atoms. Another further confirmation to our picture is the matching sign of the interchain hopping. In our model, this sign and thus the order of the irreps under which the eigenstates transform is opposite to the one presented in Ref. \cite{Sekine88}, where the roles of $A^*$ and $Z^*$ would have been reversed in the discussion. Fig. \ref{Fig:Unfold}b) shows the bands for the $D_{2d}$ structure, which are remarkably similar to those in Fig. \ref{Fig:Unfold}a), except for the fact that the faint replica of the $A^*$ at $Z^*$ is now stronger, as expected from the folding pattern discussed in the previous section. This increase in spectral weight can be taken as a measure of the extra distortion that makes three inequivalent Ta sites.

This unfolded band structure can now be compared with the one measured in ARPES experiments. It is well known that in the presence of a very weak periodic modulation, the ARPES spectral weight remains brightest in the original Brillouin Zone (BZ) before the perturbation \cite{Voit00,Rossnagel11}, with backfolded features proportional to the weak modulation. Refs. \cite{Vescoli00,Grioni01} measured the band structure dispersion of (NbSe$_4$)$_3$I in the out of plane direction, identifying a bright band edge around -0.7 eV at the $Z^* = \tfrac{2\pi}{c}$ point. Refs. \cite{Vescoli00,Grioni01} also identified that this bright spectral feature, which is found only on the second BZ of the original lattice structure, appears because of the approximate translation symmetry of $c/2$. However, it is clear from our calculation that this feature does not represent a true band edge, as it further disperses upward in the in-plane direction. Following this band to the $A^*$ point we find the true valence band edge, which is also weakly folded at $\Gamma$. The reason why it appears at $A^*$ is the presence of the approximate in-plane translation $a(\tfrac{1}{2},\tfrac{1}{2},0)$. An ARPES experiment probing the $A^*$ point should therefore identify the true band edge and provide a reliable bound for the gap. 

\section{Optical conductivity} \label{OptCon}

The second implication of the unfolded model applies to the optical conductivity, which measures optical transitions at zero momentum. To a very good approximation, these optical transitions should be considered in the unfolded band structure, where the gap is indirect. The true direct gap has very low spectral weight and should contribute little to the conductivity. The optical conductivity will only be finite for energies above the direct gap, which connects the conduction and valence bands at $Z^*$ (in the true unit cell, it connects the second highest valence band to the conduction band). In fact, to a very good approximation, there are optical transitions at around 0.5 eV for all momenta in the plane $k_z = \pi$, which should give rise to a very strong peak in the optical conductivity at that energy. 

\begin{figure}[h]
    \centering
    \hspace*{-0.35cm}
    \includegraphics[width=0.54\textwidth]{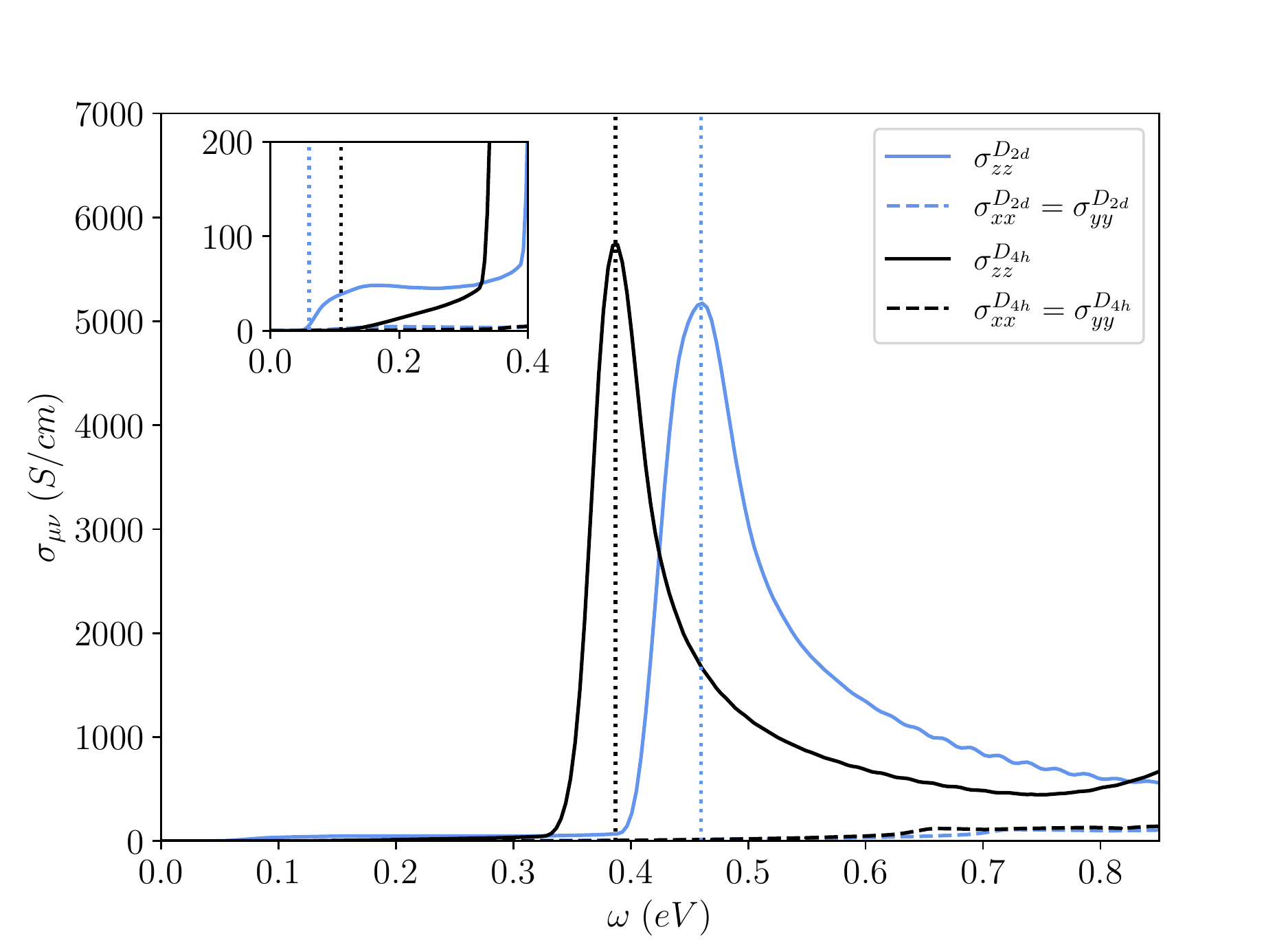}
    \caption{Optical conductivity of (TaSe$_4$)$_3$I for phases $D_{4h}$ (black) and $D_{2d}$ (blue). The locations of the peaks are marked with vertical dotted lines. The inset zooms in on the low frequency part, where the true ab-initio gap is marked with vertical dotted lines as well. }
    \label{Fig:Optical}
\end{figure}

To confirm these predictions, we have performed a calculation of the optical conductivity given by the following equation:\cite{dfg20,Sipe93}
\begin{align}
    \sigma^{ab}(\omega) = \frac{\pi e^2}{\hbar \omega V}\sum_{\mathbf{k},n,m}\text{Re}\left[ v_{nm}^av_{mn}^b \right]\delta(E_{\mathbf{k},mn}-\omega),
\label{eq:optcon}
\end{align}
where $\omega$ is the frequency of the incident electric field, $e$ is the electron charge, and $v_{nm}^a = \bra{n}\partial_a H\ket{m}$. The calculations were performed on a $91\times 91 \times 71$  $k$-mesh employing a Wannier-interpolated Hamiltonian obtained using {\tt Wannier90}\cite{Wannier90}. The results for both $D_{4h}$ and $D_{2d}$ phases are presented in Fig.~\ref{Fig:Optical}. We observe a very large peak in $\sigma_{zz}$ for both $D_{4h}$ (at $0.39$ eV) and $D_{2d}$ (at $0.46$ eV) phases, consistent with the large peak observed in experiments~\cite{Vescoli00}. As explained before, the location of this peak is no indication of the true gap of the system. We also observe that the peak is absent in $\sigma_{xx}$ as observed in experiments, and that $\sigma_{xx} \ll \sigma_{zz}$ in general because the dispersion is much weaker in the in-plane direction. 

To ascertain whether the true gap can be detected in optics, the inset of Fig.~\ref{Fig:Optical} shows a zoom of the low energy conductivity, showing that a small rise of about $75\; S/cm$ is observed for the $\sigma_{zz}$ component for both phases at the predicted ab-initio gap. Nonetheless, this small increase will hardly be appreciable on experiments, especially because the high energy peak is actually broader due to lifetime effects (which are not included in our calculation), and these low energy features have to be detected on the rising tail of this high energy peak. It is also interesting to note that the low energy rise is actually stronger for the $D_{2h}$ phase. This occurs because the folding perturbation is stronger in this phase, so there is more spectral weight for the transitions near $\Gamma$ which results in a stronger response.

\section{Doping holes vs electrons} \label{vanhove}

The recent experiment in Ref.~\cite{Bera21} synthesized a $D_{2d}$ structure of (TaSe$_4$)$_3$I which is metallic for a wide range of temperatures, and displays coexisting ferromagnetism and superconductivity a low T. This is inconsistent with the insulating band structure obtained for this compound. Interestingly, $D_{2d}$ (TaSe$_4$)$_3$I shows the smallest of all band gaps with the PBE functional, so a first possibility to explain this could be that samples in Ref.~\cite{Bera21} have somehow closed the gap via some unknown distortion. Nevertheless, the gap is larger with the more reliable mBJ functional, and our own calculations suggest that the gap is robust to small changes in lattice constant or structure, so we believe a more plausible explanation is simply extrinsic, unintentional doping of the samples. This could have occurred because the samples did show some non-stochiometry \cite{Bera21}. If this is the case, one might ask whether there is any difference between doping electrons or holes, in particular regarding the ferromagnetic and superconducting instabilities. 

\begin{figure}[t]
    \centering
    \hspace*{-0.5 cm}
    \includegraphics[width=0.5\textwidth]{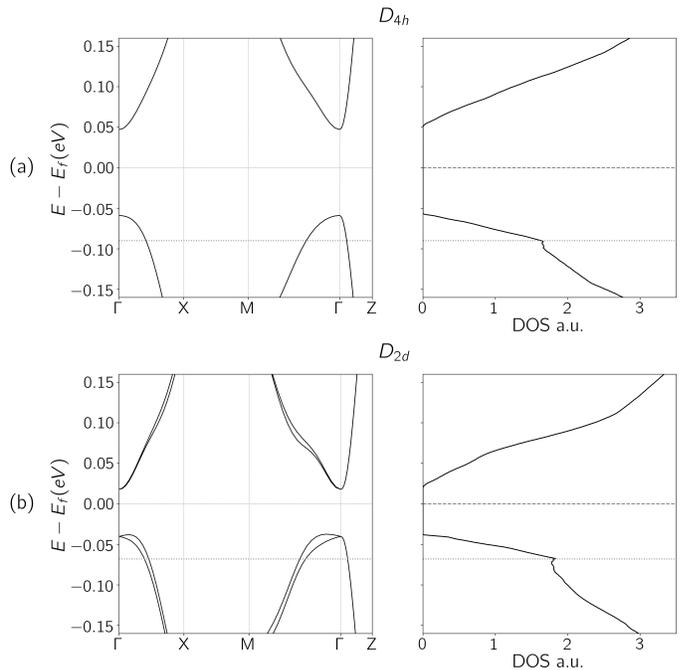}
    \caption{Ab-initio low energy band structure (left) and total DOS (right) obtained using PBE approximation for (a) Ta $D_{4h}$ phase and (b) Ta $D_{2d}$ phase. Note a Van Hove singularity for holes marked with a dotted line in both cases.}
    \label{Fig:pDOS_all}
\end{figure}

\begin{figure}[t]
    \centering
    \hspace*{-0.1 cm}
    \includegraphics[width=0.52\textwidth]{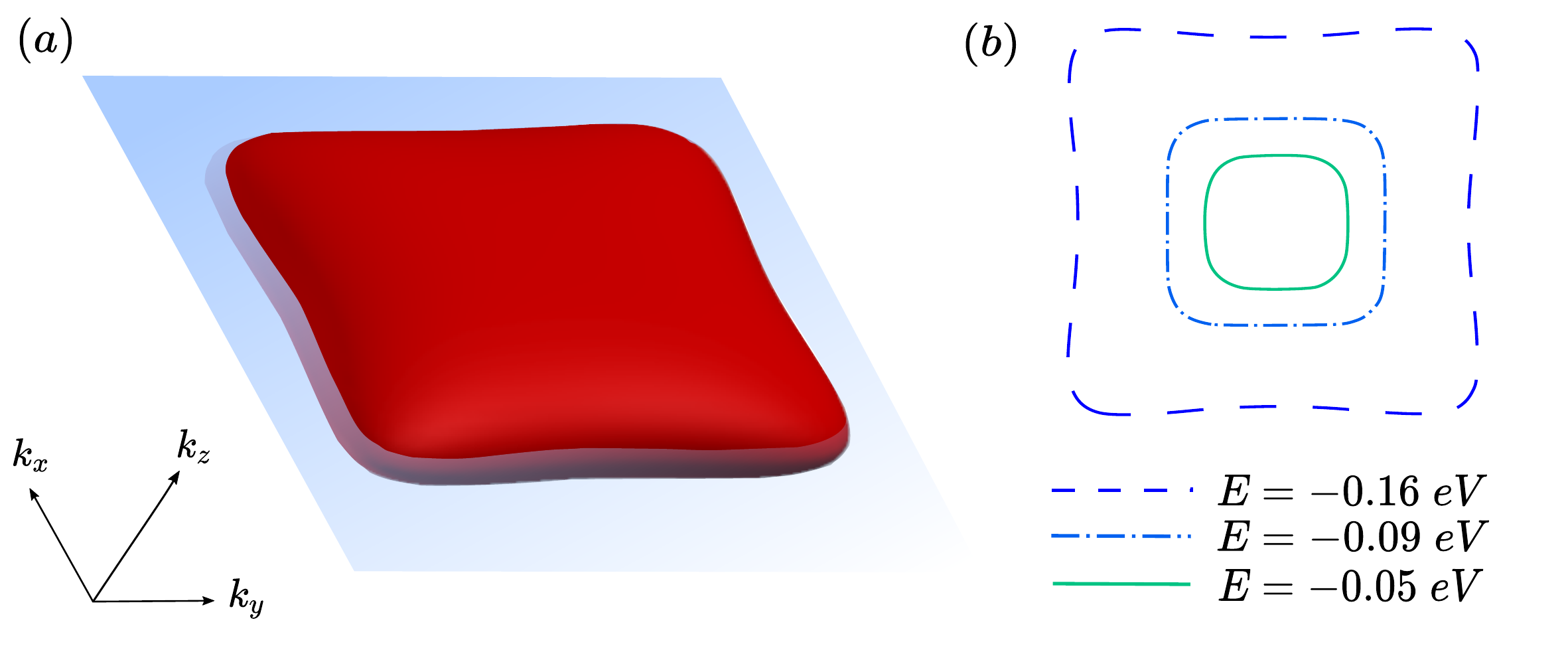}

    \caption{(a) Fermi surface for phase $D_{4h}$ at $E=-0.16$ eV. The plane $k_z=0$ is shown in blue. (b) $k_z=0$ section of the Fermi surface for phase $D_{4h}$ at different energies with respect to the Van Hove singularity: slightly below (dashed), at the singularity (dotted-dashed) and slightly above (straight).}
    \label{Fig:FermiS}
\end{figure}

To answer these questions, we have computed the DOS for the (TaSe$_4$)$_3$I $D_{4h}$ and $D_{2d}$ structures, shown in Fig. \ref{Fig:pDOS_all}. This reveals that while doping with electrons leads to a rather smooth increase of the carriers, doping with holes leads to a faster raise up to a Van Hove singularity at $E_{VH}^{D_{2d}} \approx -0.07$ eV and $E_{VH}^{D_{4h}} \approx -0.09$~eV for the two phases. This singularity emerges due to a change of shape of the Fermi surface from convex to concave as the energy is lowered, as shown in Fig. \ref{Fig:FermiS}. 

The presence of this Van Hove singularity in the valence band suggests that a ferromagnetic instability could be triggered by a small amount of hole doping, as it is predicted to occur in GaSe~\cite{Cao15} and similar monochalcogenides~\cite{Iordanidou18,Meng20,Stepanov21}. Van Hove singularities in general display different types of instabilities due to the enhanced density of states~\cite{Gonzalez00}, and their generic phase diagrams include both ferromagnetic states and unconventional pairing~\cite{Qin19,Trott20} mediated by the repulsive Coulomb interactions~\cite{Fay80}. While more work is needed to measure the amount and type of doping, as well as the Fermi surface shape, we believe the existence of this Van Hove singularity is a unique feature of the hole doped system which we conjecture will play a role in the low temperature instabilities.  

\section{Discussion}

By providing the first detailed band structure characterization of the (MSe$_4$)$_3$I compounds, our analysis has revealed that despite having a very complex lattice structure with 64 atoms in the unit cell, their low energy electronic structure is actually very simple. 
It can be understood in terms of a quasi 1D effective model of $d_{z^2}$ orbitals in a chain with a three site unit cell, and a trimerization that gives rise to a gap~\cite{Gressier84b,Gressier85b}. Our detailed characterization has explained a number of puzzles in the literature, and will serve to properly interpret future experiments in this class of materials. 

First, we have provided a quantitative prediction for the transport gaps of the different structures, which are broadly consistent with experimental observations. Our results show that the gap magnitude is not only correlated with the amount of trimerization in the different structures, but also with the unit cell volume. In the future, our predictions will also be relevant confirm the proposed distinction between type I and II samples in transport experiments, and to clarify the transport properties of the $S_4$ low temperature phase. 

Second, our work provides a clear resolution to the discrepancy between the gaps reported in transport vs those in ARPES and optical conductivity. These experiments were carried out in the high temperature $D_{4h}$ phase, whose band structure is essentially that of an indirect gap semiconductor with valence band at $A^*$ and conduction bands at $Z^*$. A very weak modulation folds both bands to $\Gamma$, but observing them in ARPES or probing an optical transition between them is extremely hard due to the very low spectral weight, proportional to the weak modulation. The large gaps quoted in ARPES and optical conductivity actually correspond to the direct gap in the unfolded bands, which is much larger than the true gap. Our predictions can be readily tested in new ARPES experiments probing the $A^*$ point. In addition, future low temperature measurements in the $D_{2d}$ phase might be more sensitive to detect the true gap in optical conductivity and ARPES, as we have shown. 

Finally, our work calls for more studies to understand the origin of the metallic behaviour of $D_{2d}$ (TaSe$_4$)$_3$I studied in Ref. \onlinecite{Bera21}. Both ARPES and optical conductivity will be useful to quantify the existence of any extrinsic doping. In addition, if the doping is hole-like, ARPES experiments can directly map the Fermi surface and confirm the existence of a low energy Van Hove singularity, which will be relevant to understand the magnetic and superconducting instabilities. We hope our work will motivate further studies on the subject to explain this unusual coexistence.  

\section{Acknowledgements}
We acknowledge N. Schroeter for clarifying discussions on ARPES experiments and J. Iba\~{n}ez-Azpiroz for useful discussions  and computational resources. F. J. acknowledges funding from the Spanish MCI/AEI/FEDER (grant PGC2018- 101988-B-C21). F. J. and M. G. V. acknowledge funding from the Diputacion de Gipuzkoa through Gipuzkoa Next (grant 2021-CIEN-000070-01). M.G.V. thanks support from the Spanish Ministerio de Ciencia e Innovacion (grant PID2019-109905GB-C21) and European
Research Council (ERC) grant agreement no. 101020833. M.G.V and C.F are thankful to the Deutsche Forschungs-gemeinschaft (DFG, German Research Foundation) – FOR 5249 (QUAST).

\bibliography{TSI}
\end{document}